# Continuous spectral and coupling-strength encoding with dual-gradient metasurfaces

Andreas Aigner[1], Thomas Weber[1], Alwin Wester[1], Stefan A. Maier[2,3], and Andreas Tittl[1,*]

[1]Chair in Hybrid Nanosystems, Nano-Institute Munich, Faculty of Physics, Ludwig-Maximilians-Universtität München, Königinstraße 10, 80539 München, Germany

[2]School of Physics and Astronomy, Monash University, Wellington Rd, Clayton VIC 3800, Australia

[3]The Blackett Laboratory, Department of Physics, Imperial College London, London, SW7 2AZ, United Kingdom

*Corresponding author. Email: Andreas.Tittl@physik.uni-muenchen.de

## Abstract

Enhancing and controlling light-matter interactions is crucial in nanotechnology and material science, propelling research on green energy, laser technology, and quantum cryptography. Central to enhanced light-matter coupling are two parameters: the spectral overlap between an optical cavity mode and the material's spectral features (e.g., excitonic or molecular absorption lines), and the quality factor of the cavity. Controlling both parameters simultaneously is vital, especially in complex systems requiring extensive data to uncover the numerous effects at play. However, so far, photonic approaches have focused solely on sampling a limited set of data points within this 2D parameter space.

Here we introduce a nanophotonic approach that can simultaneously and continuously encode the spectral and quality factor parameter space of light-matter interactions within a compact spatial area. Our novel dual-gradient metasurface design is composed of a 2D array of smoothly varying subwavelength nanoresonators, each supporting a unique mode. This results in 27,500 distinct modes within one array and a resonance density approaching the theoretical upper limit for metasurfaces. By applying our dual-gradient to surface-enhanced molecular sensing, we demonstrate the importance of coupling tailoring and unveil an additional coupling-based dimension of spectroscopic data. Our metasurface design paves the way for generalized light-matter coupling metasurfaces, leading to advancements in the field of photocatalysis, chemical sensing, and entangled photon generation.

# Introduction

Optical cavities have significantly advanced our ability to manipulate light-matter interactions, with various applications ranging from lasers and spectroscopic techniques to quantum information processing[1]. Especially in nanoscience, nanoresonators[2,3], the nanoscale counterparts of optical cavities, have bridged the size gap to materials like quantum dots, Van der Waals materials, and molecules. This allowed breakthroughs in photocatalysis[4,5,6], entangled photon sources[7,8], biochemical sensing[9,10], and the study of polaritons[11,12]. The interaction of light and matter in nanoresonators is fundamentally governed by two key parameters: the spectral overlap of the optical mode with the excitation of the target system (e.g., excitonic[13,14], or molecular absorption lines[15,16]) and the strength of the interaction set by the resonators' quality factor (Q-factor, defined as the resonance frequency divided by the linewidth).

In terms of spectral overlap, the fundamental goal is to simultaneously amplify and probe the material's dispersive properties. In pursuit of achieving the necessary spectral coverage, research in multiresonant nanophotonic platforms has led to concepts like plasmonic oligomers[17,18], multiresonant plasmonic surface lattice resonances[19,20], dual band perfect absorbers[21], fractal plasmonics[22,23], and nonlocal metasurfaces[24]. However, these platforms often face restrictions owing to the limited number of resonances that a single nanoresonator, or 2D arrays of such resonators, known as metasurfaces[25,26], can support. While active metasurfaces allow some degree of tunability, they typically cannot cover wide spectral ranges[12,27]. Thus, researchers have turned to using metasurfaces with spatially varying resonator geometries which allows for local adjustment of the optical response. Initially developed for non-resonant phase gradient metasurfaces[28,29], this concept has been extended to resonant systems[30,31].

Whereas the advancements for spectral coverage have been significant, precisely tuning the resonators' coupling strength remains challenging. Symmetry-protected bound states in the continuum (BICs) have recently been proven to be a promising solution[32,33]. By adjusting geometric parameters, BICs can fine-tune their resonance linewidths, controlling the strength of the light-matter interaction[34,35]. This has led to multispectral BIC-driven metasurfaces with up to 100 resonances on a single sensor platform[36,37]. While BICs are effective at probing the spectral and coupling space, they require extended arrays of identical resonators as BICs inherently are collective modes[38]. Each metasurface, consisting of at least hundreds of identical resonators, can only probe a single point in the 2D spectral-coupling parameter space. This prohibits the complete analysis of complex systems that need extensive data to uncover the numerous interactions and effects occurring within them. Further, it hinders the integration into hyperspectral optical systems and compact devices.

Despite recent successes in creating spectrally tuned plasmonic gradients for refractive index sensing[39] and dielectric gradients for higher harmonic generation[40], a comprehensive platform for studying both the coupling space and the combined 2D spectral-coupling space remains elusive.

Here, we introduce the novel concept dual-gradient metasurfaces, seamlessly spanning the 2D parameter space of resonance wavelength and coupling-strength, ideal for probing diverse light-matter coupling phenomena. First, we explore spectral gradients which offer continuously spectrally tunable resonances. Experimentally, we investigate the effect of the gradient's spectral width on the resonance performance. Strikingly, we find that for moderate widths, the performance of the gradient is equal to established monospectral metasurfaces, despite the perturbed periodicity. This finding is crucial for applications that require continuous spectral coverage without compromising resonance performance, such as on-chip spectrometers and integrated photovoltaic devices. Similarly, we present the idea of coupling (Q-factor) gradients. This concept provides a spatial mapping of light-matter interaction strength, making them ideal for studying nonlinear effects, photon generation processes, or the excitation of polaritons. Combining these advancements, we experimentally realized a dual-gradient metasurface with independently adjustable gradients, showcasing extensive and simultaneous spectral and coupling-strength coverage in a compact footprint. To demonstrate the dual gradients' capabilities, we apply it to surface-enhanced molecular sensing, where a wide spectral coverage is needed to retrieve the unique vibrational fingerprints of molecules. We not only capture the spectral fingerprint but also unveil an additional coupling-based dimension of spectroscopic data, showcasing a concentration-based dependence between the resonances' Q-factor and the detection sensitivity. The development of spectral and coupling gradients as well as their combination into dual-gradient metasurfaces marks a substantial advancement over conventional metasurface approaches. Significantly, this innovative approach opens new opportunities for research in fields like photonic computing and polaritonics, where complete analysis of the spectral and coupling parameter space is needed, and broadens the scope of potential applications, including on-chip spectrometers, sensors, photocatalytic reactors, and entangled photon sources.

## Results

We chose a BIC-driven metasurface geometry consisting of pairs of tilted amorphous silicon ellipses (**Figure S1a**) on top of an infrared-transparent calcium fluoride ($CaF_2$) substrate. This particular unit cell design is ideal for our target molecular spectroscopy application due to its strong surface-confined electromagnetic fields, low baseline reflectance, and fabrication robustness[41]. Extending the traditional approach of arranging the elliptical resonators in a two-dimensional periodic array, we introduce a multiplicative lateral scaling factor $S$ to modify the dimensions of each unit cell, creating spectral gradients (**Figure 1a**). Spectral resonance tuning is achieved by continuously varying $S$, creating an unambiguous mapping between spatial and spectral information. The photonic behavior of symmetry-protected BIC metasurfaces is fundamentally determined by the asymmetry factor $\alpha$, defined as the sine of the ellipse tilt angle $\theta$ via $\alpha = sin(\theta)$[42]. The asymmetry controls the radiative coupling of the resonance to the far-field and therefore governs the resonance's quality (Q) factor, the ratio of the resonance frequency to the resonance bandwidth (**Figure S1b,c**). By arranging unit cells with identical $S$ but varying $\alpha$ into a two-dimensional lattice, we generate a coupling gradient that provides a broad range of Q-factors within a single metasurface (**Figure 1b**).

Following extensive numerical optimization, targeted at achieving BICs on the long wavelength side of the Rayleigh limit[43], even for high asymmetries $\theta$ of up to 45° (see **Figure S1d-i** for details), we have selected a unit cell geometry with a pitch of $P_x = 2.4$ μm in the x-direction, a pitch of $P_y = 4$ μm in the y-direction, an ellipse long diameter $A$ of 2 μm, a short diameter $B$ of 1 μm, and a height $h$ of 0.75 μm. The numerical results reveal that adjusting the scaling factor from 1.0 to 1.3 along the x-axis for $\theta = 20°$ shifts the pronounced reflectance peak of the BIC resonance from 5.9 μm to 7.2 μm, underlining the distinct spatially dependent response of a spectral-gradient metasurface (**Figure 1c**). Similarly, **Figure 1d** presents numerical results for a fixed scaling factor ($S = 1$) with the tilting angle $\theta$ ranging from 0 to 45°, generating the spatially dependent response of a coupling gradient. We observe a broadening of the mode with increasing angles, following the typical Q-factor relationship for symmetry-protected BICs, in this case $Q = 1/sin^2(\theta)$.

The unique simultaneous parameter variation of the dual gradient is created by merging both resonance tuning mechanisms (**Figure 1e**), allowing the metasurface to seamlessly encode a wide range of spectral and coupling-strength information. This unprecedented capability makes dual-gradient metasurfaces ideal for probing and reading out the rich spectral and coupling-strength information found in many materials, like Van der Waals materials, quantum dots, or molecules with their unique vibrational fingerprints.

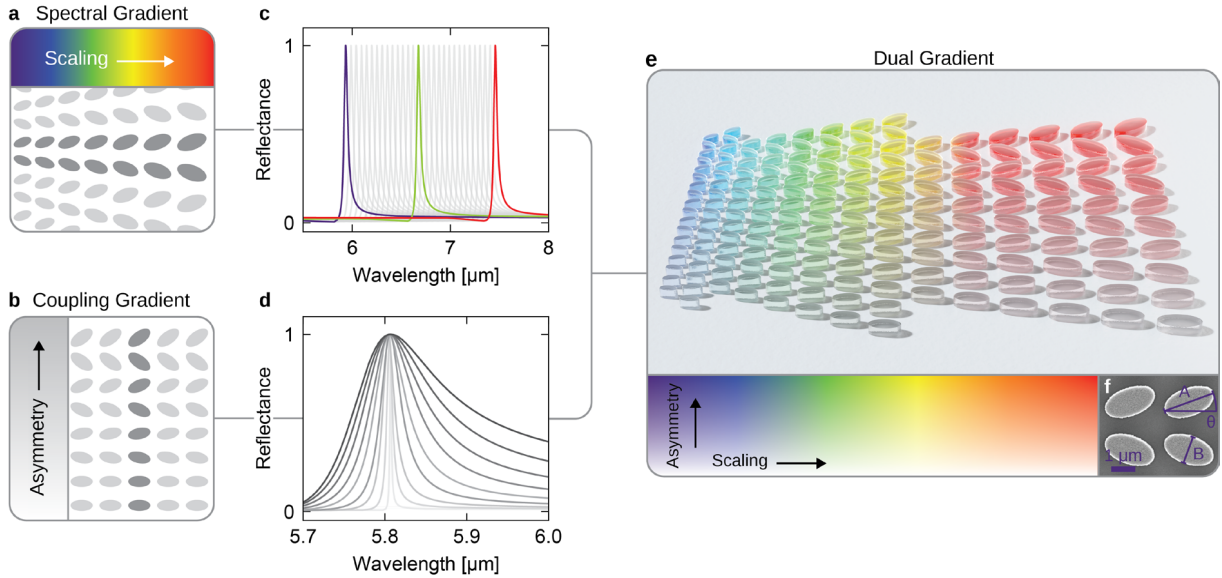

**Figure 1. Concept of dual-gradient metasurfaces combining independent spectral and coupling gradients.** (a) Schematic representation of a spectral-gradient metasurface composed of unit cells of tilted ellipse pairs. Identical resonators form chains along the vertical, with a gradual increase in scaling along the horizontal. (b) Schematic representation of a coupling-gradient metasurface. The asymmetry (in the form of the tilting angle $\theta$) varies along the vertical, altering the far-field coupling-strength and the resonance Q-factor. (c) Numerical reflectance spectra of BIC resonances with scaling factors ranging from 1.0 to 1.3, increasing incrementally by steps of 0.01. (d) Numerical reflectance spectra with a constant scaling factor ($S = 1$), and $\theta$ varying from 0° (light gray) to 45° (black) in increments of 5°. Spectral alignment of the resonances is achieved by using an additional scaling factor, discussed in Figure 3. (e) Illustration of a dual-gradient metasurface that incorporates both a spectral gradient along the horizontal and a coupling gradient along the vertical.

We initiate our experimental demonstration of gradient metasurfaces by investigating the performance of continuous spectral gradients with a fixed tilting angle $\theta = 20°$ and varying lateral scaling $S$. Using high-resolution electron-beam lithography and reactive ion etching, a spectral gradient with continuous unit cell scaling along the x-axis from $S = 1.0$ to 1.1 over a length of 600 µm was realized (scanning electron microscopy (SEM) image in **Figure 2a**, angled view in **Figure S2**). We directly compared it with three monospectral metasurfaces (150x150 µm², 35x59 unit cells) with $S = 1.01$, 1.05 and 1.09, fabricated in proximity right above the spectral gradient (optical image in **Figure 2b**). The monospectral metasurfaces were chosen so that they correspond to points in the center and close to the ends of the spectral gradient, with resonance wavelengths of 5.9 µm, 6.1 µm, and 6.3 µm. The metasurfaces were optically characterized using a multispectral imaging microscope incorporating tunable quantum cascade lasers and a 480x480 pixel imaging detector (**Figure S3a**), allowing us to record snapshots of the reflectance signal at different wavelengths.

At each of the three wavelengths given above, one of the monospectral metasurfaces exhibits a high reflectance amplitude, indicating resonant nanostructures. The spectral gradient, however, shows pronounced reflectance zones (cross-sections for 6.1 μm in **Figure S4**) across all three wavelengths, demonstrating resonant behavior throughout (**Figure 2c**). **Supplementary Video 1** illustrates this unique resonant behavior, where each frame represents the reflectance response for a specific wavelength. In striking contrast to the monospectral metasurfaces, the peak reflectance map across all wavelengths shows a consistently high reflectance signal over the whole spectral gradient, indicating efficient and continuous resonance coverage (**Figure S3b**, rotated polarization in **Figure S5a**). This advantageous behavior is further highlighted by plotting the extracted resonance wavelengths $\lambda_{\text{res}}$ for each image pixel (**Figure 2d**), where the spectral gradient reveals a smooth transition from 5.9 to 6.4 μm, compared to the solid colors associated with the discrete metasurface pixels. The reduced reflectance amplitude in the upper portion of both the gradient and the monospectral metasurfaces can be attributed to the instrument response of our optical spectroscopy setup, where rotating the sample by 180° shifts this effect to the opposite side (**Figure S5b,c**).

To further assess the performance of the spectral gradient, we extract normalized reflectance spectra from equally distributed points along the gradient's x-axis (**Figure 2e**, gray lines) and compare them to the monospectral metasurfaces, normalized to the reflectance amplitudes of the gradient at the corresponding resonance positions (**Figure 2e**, colored lines). Strikingly, the spectra taken from the gradient with $\Delta S = 0.1$ show nearly identical reflectance amplitudes compared to the monospectral metasurfaces. The insight that a perturbation of the periodicity has no negative effect as long as it remains below a moderate value is a new insight into BICs, and is crucial for applications requiring both continuous spectral coverage and maximal resonance performance.

As a next step, we expand our analysis beyond the $\Delta S = 0.1$ gradient and investigate gradients of the same lateral size but with a varying spectral range and thus varying steepness. In total, we fabricated and analyzed eight gradients with spectral ranges from 225 nm to 1910 nm ($\Delta S = 0.05 - 0.4$). **Figure 2g** displays the average maximum reflectance amplitude for the different gradients, with their spectral coverage visualized by the bar length. We find a considerable decrease in resonance amplitude with increasing spectral coverage, dropping from an average reflectance amplitude of 0.52 for the 225 nm coverage case to 0.15 for the 1910 nm coverage case. To quantify this correlated behavior, we introduce the scaling increment $\varepsilon_S = \Delta S \cdot P / \Delta L$, calculated from the maximum scaling factor variation $\Delta S$, unit cell periodicity $P$, and spatial extent $\Delta L$ of the gradient metasurface. This metric offers an intuitive way to characterize the behavior of the gradient, since it represents the change in scaling factor between two neighboring unit cells. Extracted average reflectance amplitudes and Q-factors for different tilting angles $\theta$ (10°, 20°, and 30°) are shown in **Figure 2h** as a function of $\varepsilon_S$. For comparison, the performance of the monospectral metasurfaces for $\theta = 20°$ is shown as dashed lines. Although the average resonance amplitude and Q-factor decrease as $\varepsilon_S$ increases, the spectral gradients can match the performance of the monospectral metasurfaces when $\varepsilon_S \leq 0.5 \cdot 10^{-3}$, as highlighted for the case of $\theta = 20°$ (**Figure 2h**). Crucially, even in the parameter range where reflectance and Q-factor fall below the monospectral case ($\varepsilon_S \geq 0.5 \cdot 10^{-3}$), spectral gradients can still provide considerable benefits, such as hyperspectral operation, while maintaining sufficient optical performance. These advancements pave the way for novel chip-integrated solutions and ultra-compact spectroscopy devices.

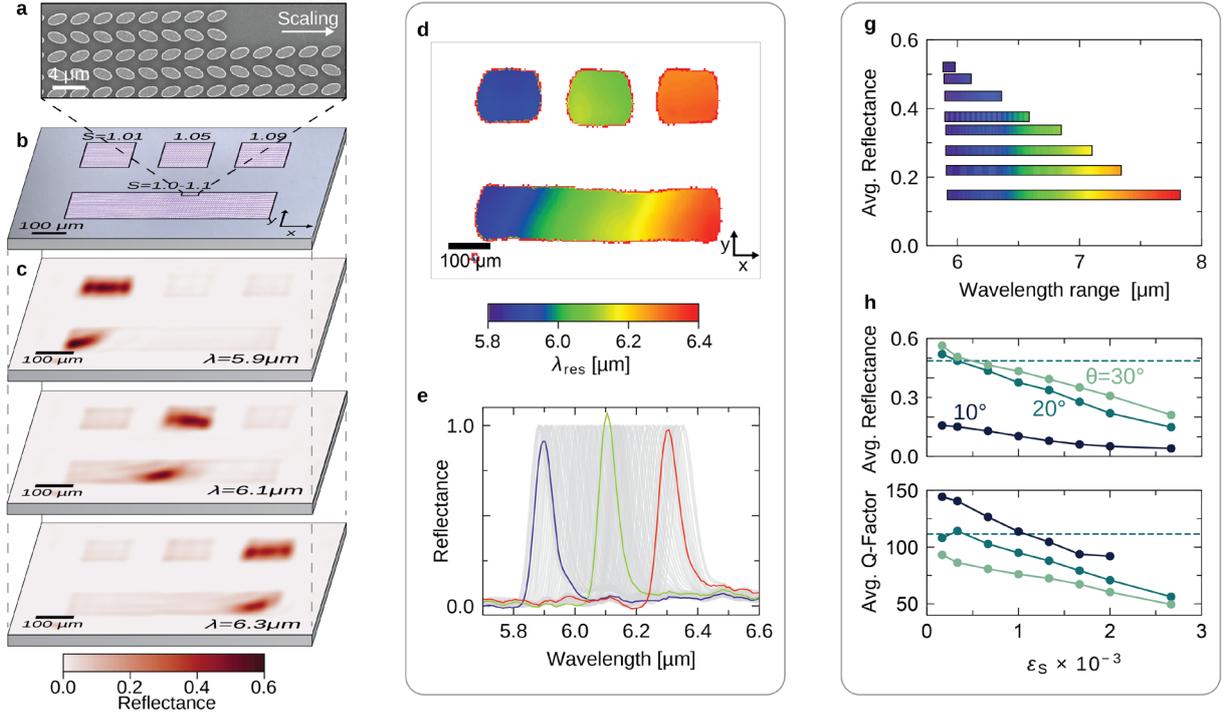

**Figure 2: Spectral-gradient metasurface.** (a) Scanning electron microscopy (SEM) image of a spectral gradient with the scaling along the x-axis. (b) Optical image of the three monospectral metasurfaces with scaling factors $S = 1.01, 1.05$, and $1.09$ from left to right and below the spectral gradient with $S = 1.0 - 1.1$. (c) Reflectance snapshots of the metasurfaces shown in (b), taken at wavelengths of 5.9, 6.1, and 6.3 µm. (d) Color-coded resonance wavelengths $\lambda_{res}$ for each detector pixel. Non-resonant pixels are shown in white. (e) Normalized reflectance spectra taken from the spectral gradient along the x-axis, shown in gray. The colored spectra correspond to the three monospectral metasurfaces, each normalized to the gradient's maximum reflectance at their respective spectral position. (g) Average reflectance amplitude and wavelength range for gradients with $\theta = 20°$ of different gradient steepness. (h) Average reflectance amplitude and Q-factor plotted against different scaling increments $\varepsilon_S$ for gradients with $\theta = 10, 20$, and $30°$. The horizontal dashed line represents the values for the monospectral metasurfaces with $\theta = 20°$.

Following the successful demonstration of high-performance spectral gradients, we now focus on spatially encoding the radiative losses of BIC-driven metasurfaces. As introduced in **Figure 1b**, tuning of the ellipse opening angle $\theta$ can provide resonances with a wide range of Q-factors, following the characteristic inverse square relationship $Q = 1/\alpha^2 = 1/sin^2(\theta)$ (**Figure S1b**)[42]. This precise resonance control allows for tailored interactions between the resonant mode and surrounding materials since the Q-factor directly influences the local electromagnetic field enhancement $FE$ via $FE^2 \propto Q^{44}$ (**Supplementary Note 1**).

Leveraging the concept of coupling gradients introduced above, we fabricated metasurfaces with continuously increasing values of $\theta$ ranging from 0° to 45° (**Figure 3a**). The direction of the $\theta$ variation is chosen perpendicular to the excitation polarization, since this configuration provides better optical performance for spectral gradients (**Supplementary Note 2, Figure S6**). The dimensions of the gradient are 650x150 µm². Experimental reflectance spectra in **Figure 3b** show overall increasing amplitudes for increasing $\theta$, which we attribute to the reduced susceptibility of lower-Q resonances to intrinsic material losses and fabrication defects. We attribute the decrease in amplitude for high $\theta$ to the quenching effect of the Rayleigh limit. Using temporal coupled mode theory (TCMT, **Supplementary Note 1, and Figure S7** for a schematic illustration), we fit the reflectance spectra of each pixel of our dataset to extract the characteristic resonance parameters (reflectance amplitude, resonance wavelength $\lambda_{res}$, Q-factor). The resulting Q-factor map in **Figure 3c** (maximal reflectance map in **Figure S8**) aligns with our numerical design, showing a decrease in Q-factor with increasing $\theta$. However, when changing the asymmetry, $\lambda_{res}$ does not remain constant but undergoes a spectral shift of around 500 nm (**Figure 3d**). This crosstalk between $Q$ and $\lambda_{res}$ is detrimental for applications, as it hinders the consistent spectral overlap with dispersive media like molecules and their vibration lines.

To overcome this challenge and facilitate the accurate implementation of dual-gradient metasurfaces, we introduce the concept of spectrally-aligned coupling gradients using an additional ellipse scaling factor $ES$. Instead of modifying all lateral unit cell parameters, $ES$ solely alters the ellipses' dimensions, $A$ and $B$, while keeping $P_x$ and $P_y$ constant, allowing for local adjustments of the resonance wavelength without breaking the overall gradient metasurface pattern. Using extensive numerical simulations to determine $ES$ values that offset the spectral shifts caused by $\theta$ (**Figure 3e**, **Figure S9**), we then fabricated a spectrally-aligned coupling gradient as shown in the sketch and SEM images at $\theta = 0°$ and $45°$ in **Figure 3f**.

Experimental reflectance spectra for different asymmetries confirm the near-perfect spectral alignment of the resonances (**Figure 3g**, maximal reflectance map in **Figure S8**). Additionally, the reflectance amplitude is significantly improved compared to the initial gradient, especially for large asymmetries (**Figure 3h**). Likewise, the spectrally-aligned coupling gradient delivers consistently higher values of the Q-factor for all tilting angles $\theta$ (**Figure 3i,j**), and the resonance wavelength remains nearly constant at 5.85 μm, as shown in the resonance wavelength map in **Figure 3k** and in the comparison plot in **Figure 3l**. **Supplementary Video 2** directly compares both coupling gradients frame by frame.

It is crucial to note that the relationship $FE^2 \propto Q$ holds true only in lossless systems (**Supplementary Note 1**). In our silicon resonators, while material (intrinsic) losses are minimal, scattering losses are a significant factor. These losses, arising from various sources such as surface roughness, variations in resonator size, and imperfectly collimated excitation light, lead to discrepancies between our numerical and experimental findings. Specifically, this is evident in the deviation observed between the numerical results shown in **Figure 1d**, where the reflectance amplitude is equal to 1 across all Q-factors, and the experimental results in **Figure 3g**, where the reflectance amplitude is notably reduced, especially at higher Q-factors. Accounting for these losses, the peak field enhancement in our excited gradient is achieved at $\theta = 9°$ (**Figure S10**). Beyond this point, the coupling gradient demonstrates a continuous and smooth tuning of coupling-strength and a linearly decreasing $FE^2$ with the Q-factor (**Supplementary Note 1**). Future applications of our principle could potentially shift the point of highest field enhancement to smaller asymmetries, thereby aligning more closely with the ideal scenario. **Supplementary Note 3** discusses the influence of scattering loss on the position of highest field enhancement in more detail.

Through the ability to continuously scale the resonance linewidth, and thus the local field enhancement while maintaining its resonance frequency, the spectrally-aligned coupling gradient holds great promise for various applications, from photocatalysis and higher harmonics generation to the study of resonator-coupled polaritons. Moreover, it enables the full decoupling of spectral and coupling-strength tuning, providing the crucial prerequisite for realizing a dual-gradient metasurface that encompasses the entire 2D parameter space defined by spectral and coupling-strength information.

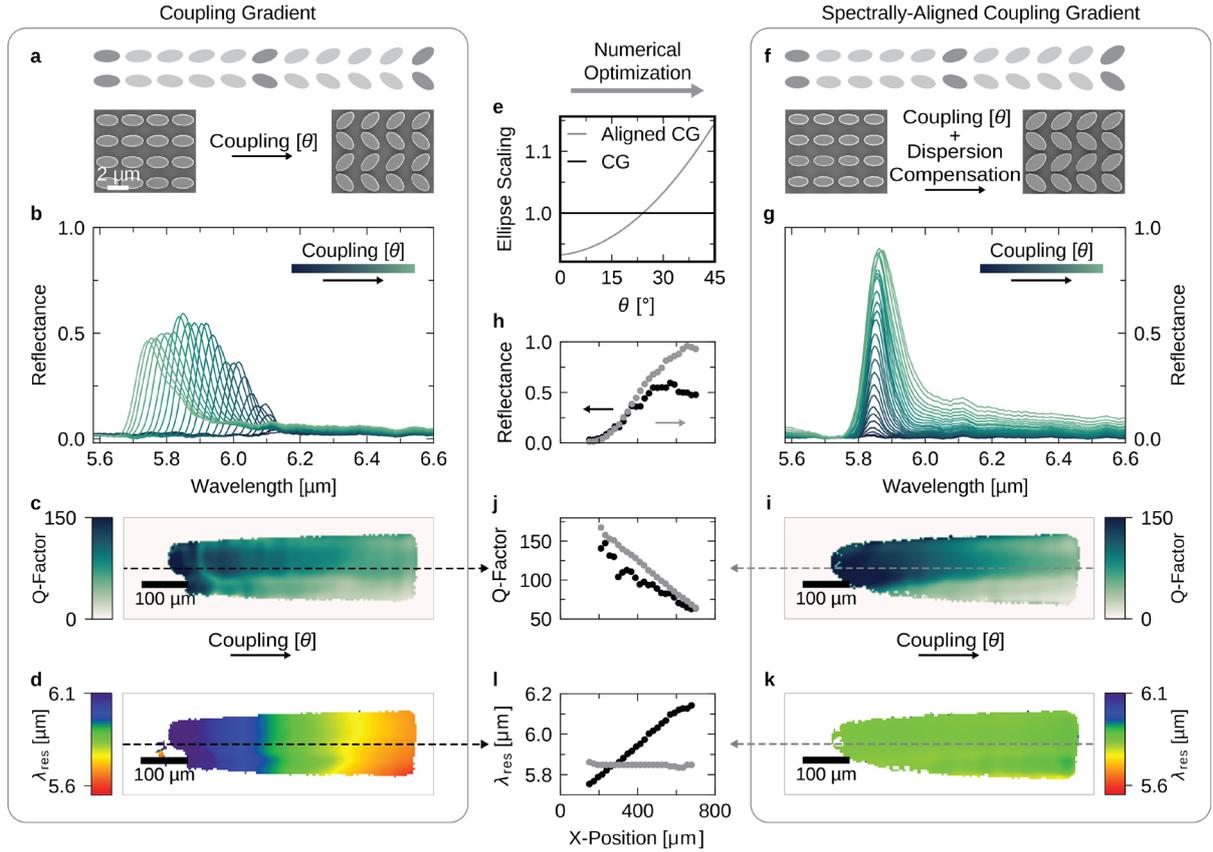

**Figure 3: Coupling-gradient metasurface.** (a) Illustration of the coupling gradient oriented perpendicular to the excitation polarization, accompanied by SEM images of unit cells from the gradient's start and end. (b) Reflectance spectra from the gradients captured along the x-axis. (c) Q-factor map derived from TCMT for the coupling gradient and (d) its associated resonance frequency map. (e) Depiction of ellipse scaling $ES$ dependent on $\theta$ for the coupling gradient (black) and the spectrally-aligned coupling gradient (gray). (f) Illustration of the spectrally-aligned gradient and SEM images of the gradient's start and end. (g) Reflectance spectra from the spectrally-aligned coupling gradient, analogous to (b). (h) Reflectance comparison of the two gradients. (i) Q-factor map for the spectrally-aligned coupling gradient and (j) a comparison between both gradients taken along the dashed lines in (c) and (i). (k) Resonance wavelength map for the spectrally-aligned gradient and (l) a comparison of both gradients taken along the dashed lines in (d) and (k).

A dual-gradient metasurface (**Figure 4a**) is realized by building on the insights gained from the spectral and coupling gradients introduced above. Continuous spectral tuning is applied along the short axis ($P_x$) of the unit cell, enabling the effective excitation of collective dipoles as discussed in **Figure S6**. Conversely, the tilting angle $\theta$ is scaled along the long axis of the unit cell ($P_y$). We fabricated a dual-gradient measuring 600x450 µm², with a spectral scaling factor $S$ ranging from 0.95 to 1.25 ($\Delta S = 0.3$) and $\theta$ from 0 to 45°. The optical performance of the dual gradient is illustrated by taking a single-wavelength reflectance image at 6.25 µm (**Figure 4b**, additional wavelengths in **Figure S11**, and **Supplementary Video 3**), showcasing a narrow vertical strip indicating excellent spectral selectivity. The vertical strip increases in width for higher asymmetries indicating precise control over the coupling-strength.

**Figure 4c** shows the extracted resonance wavelength for each pixel, obtained through TCMT modeling, demonstrating a linear and continuous spectral coverage from 5.6 to 7.2 µm (**Figure S12a,b**). Notably, the resonance wavelength is unperturbed by different values of $\theta$, as enabled by the spectral alignment procedure. **Figure 4d** shows the coupling-strength encoding, with Q-factors decreasing continuously from approximately 130 to 20 along the y-axis as $\theta$ increases (see **Figure S12c,d** for a cut along the y-axis). By combining the wavelength and coupling-strength encodings highlighted in **Figures 4c** and **4d**, we successfully demonstrate the first dual-gradient metasurface that continuously maps both the wavelength and coupling parameter space in a unified nanophotonic system.

Delving deeper into the implications of dual gradients, we introduce the density of resonances $\rho_{res}$ as a measure for the amount of information encoded within a metasurfaces. To ensure a wavelength independent metric, we define $\rho_{res}$ as the number of resonances $N_{res}$ divided by the number of unit cells $N_{unit\,cell}$ as $\rho_{res} = N_{res}/N_{unit\,cell}$ (see **Supplementary Note 4** for more details). Conventional monospectral metasurfaces, as illustrated in **Figure 4e,** typically feature only a single resonant mode ($N_{res} = 1$). Consequently, $\rho_{res}$ is low, e.g. $\rho_{res} = 9.1 \cdot 10^{-4}$ for a conventional 100x100 µm² metasurface[36]. While a shrinkage of the footprint will increase $\rho_{res}$, there is a fundamental limit to the minimum pixel size in order to sustain high-Q modes[38]. Conversely, in our spectral and coupling gradients, only 1D chains of resonators share identical geometrical parameters, resulting in a higher $\rho_{res}$ values of $2.9 \cdot 10^{-2}$ and $2.7 \cdot 10^{-2}$, respectively. The dual gradients, as illustrated in **Figure 4e**, show an even higher resonance density, with each unit cell being unique within the gradient, therefore encoding a distinct point in the 2D spectral-coupling parameter space. This pushes $\rho_{res}$ to its theoretical maximum of 1, and leads to a remarkable total number of 22,800 resonances (with 27,500 resonances due to a smaller unit cells depicted in **Figure 5**) across the dual gradient. Such a high number of resonances and $\rho_{res} = 1$ is unprecedented even for state-of-the-art metasurface designs. This breakthrough is evident in the comparative analysis presented in **Figure 4f**, where our dual-gradient design surpasses previous metasurface implementations, outperforming $\rho_{res}$ in plasmonic gratings by a factor of 50, and exceeding dielectric metasurfaces by at least two orders of magnitude, as detailed in **Supplementary Note 5** and **Table S1**. It is important to note that in our methodology, $N_{res}$ is considered the theoretical maximum number of supported resonances. However, practical limitations related to spatial and spectral resolution may reduce the actual number of distinguishable modes.

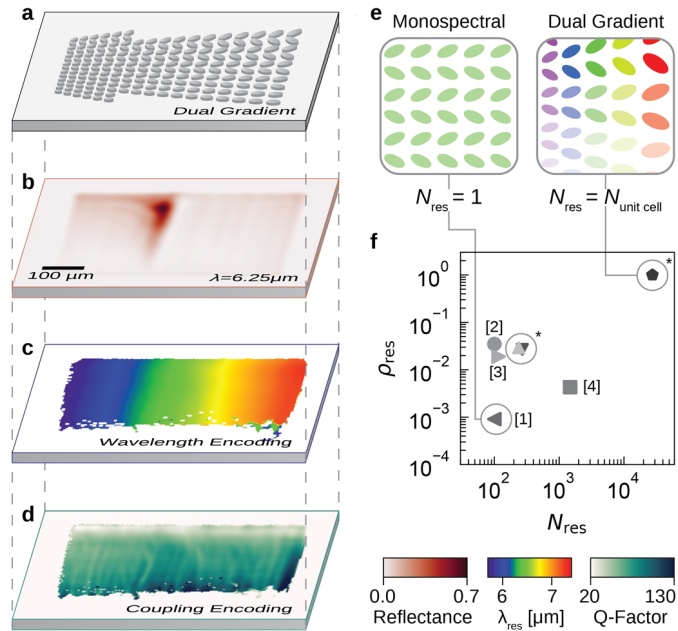

**Figure 4: Dual-gradient metasurfaces and resonance density.** (a) Sketch of a dual-gradient metasurface with the spectral gradient along the x-axis and the spectrally-aligned coupling gradient along the y-axis. (b) Single wavelength snapshot of the dual gradient with $S = 0.95 - 1.25$ and $\theta = 0 - 45°$ at 6.25 µm. (c) Resonance wavelength map of the dual gradient showing continuous wavelength encoding along the x-axis. (d) Q-factor map

of the dual gradient with decreasing values along the y-axis. (e) Illustrative comparison of distinct resonances within conventional monospectral metasurfaces and a dual-gradient metasurface. In the monospectral metasurface, all unit cells are identical, resulting in a single supported resonance ($N_{res} = 1$). In contrast, each unit cell within the dual-gradient metasurface is unique, leading to $N_{res}$ being equal to the total number of unit cells ($N_{unit\ cell}$). (f) A log-log plot of $N_{res}$ against the resonance density ($\rho_{res}$) is presented. Comparative works are marked with numbers: Pixelated sensors [1][36], radial BICs [2][45], trapped rainbows [3][46], and spectral-gradient metasurfaces [4][40]. Our work is highlighted with an asterisk (*), showcasing the spectral gradient and the coupling gradient at $\rho_{res} = 2.9 \cdot 10^{-2}$ and $2.7 \cdot 10^{-2}$ respectively. The dual gradient is positioned at the top right, with $\rho_{res} = 1$.

The dual gradient captures generalized information about light-matter coupling processes and consequently has potential applications in a wide variety of fields ranging from polaritonic coupling to quantum light emission and biochemical sensing. Here, we will focus on utilizing dual gradients to obtain new functionalities for the surface-enhanced infrared absorption spectroscopy (SEIRAS) of molecular systems[15,47]. In addition to retrieving the conventional spectral fingerprint of the molecules, the integration of the coupling parameter into the gradient metasurface unlocks a new dimension of spectroscopy data, which allows to always achieve optimal sensitivities and can be correlated with intrinsic analyte properties such as concentration. To reveal the relationship between analyte concentration and optimal sensing performance, we conducted a series of measurements on a dual-gradient metasurface ($S = 0.95 - 1.1$, $\theta = 0 - 45°$, **Figure S13**, 27,500 distinct resonances) with varying amounts of molecular analyte. We chose PMMA as the analyte because of its widespread use as a sensor benchmark[48,49,50] and the straightforward control over the layer thickness through spin-coating of solutions with varying concentrations. Seven concentrations of PMMA dissolved in anisole ranging from 4% to 0.05% were prepared and spin-coated onto the metasurface at 3000 rpm, resulting in layer thicknesses of approximately 200 (4%) to 1.5 nm (0.05%).

**Figure 5a** illustrates the varying thicknesses of the PMMA coating on top of the resonators; the uncoated structure is on the far left, followed by structures coated with decreasing concentrations of 1%, 0.2%, and 0.05% PMMA solution. Maximum reflectance images of the dual gradient with the coatings from **Figure 5a** are presented in **Figure 5b**. The molecular fingerprint of PMMA can be clearly resolved by eye for the higher concentrations, where it appears as an area of reduced maximal reflectance within the left third of the dual gradient. **Figure S14** further illustrates the molecular fingerprint within the Q-factor map for a 1% concentration. By calculating the relative absorbance of the analyte on a pixel-by-pixel basis via $A = -\log(R_C/R_0)$, where $R_C$ and $R_0$ represent the maximum reflectance of the coated and uncoated gradient metasurface, respectively, the molecular fingerprint of PMMA becomes evident for all concentrations (**Figure 5c**, wavelength by wavelength comparison in **Supplementary Video 4 & 5**).

While a decrease in overall reflectance modulation with the amount of analyte is expected, we observe an intriguing shift of the maximum relative absorbance along both the spectral and coupling-strength axes for different concentrations. To quantify this effect, we focus on a subsection of the full gradient marked by dashed black lines in **Figure 5c** and extract the 100 pixels with the highest relative absorbance for each concentration (**Figure 5d**). These pixels, which represent areas with highest sensitivity for a specific concentration, are localized in distinct and mostly connected clusters, with only few outliers caused by slight misalignments when overlaying the reflectance maxima of the coated and uncoated gradients. As the analyte concentration decreases, the spectral redshift caused by the analyte's refractive index decreases, and the highest sensitivity pixels move towards larger scaling factors, compensating for the shift. Although wavelength-based resonance shifts are not specific for molecular detection, refractometric detection is widely used for photonic sensors. There is a diverse field of research focusing on sensors that employ waveguides, plasmonic nanoparticles, and metasurfaces, particularly BIC metasurfaces due to their high Q-factors and thus high sensitivity. While refractive index sensing excels in quantifying the concentration of analytes, SEIRAS provides the needed specificity to identify the distinct molecular type. This gives the dual gradient two complementary sources of information.

In addition to this resonance wavelength-induced shift in **Figure 5d**, a distinct trend towards lower asymmetries (higher Q-factors) is observed as the concentration decreases. This phenomenon is particularly intriguing as it is not observable in conventional sensor approaches. To better illustrate this trend, **Figure 5e** depicts the mean values for all concentrations as ellipses, with diameters equal to two standard deviations along the x- and y-axis, with the linear fit line clearly showing a tendency towards higher Q-factors at lower concentrations. These findings emphasize the critical role of the coupling parameter in reaching maximal sensitivity. Our dual-gradient metasurface, with its full coverage of the coupling space, consistently operates at peak efficiency, unaffected by variations in analyte concentration or the presence of loss-inducing materials, such as solvents like water.

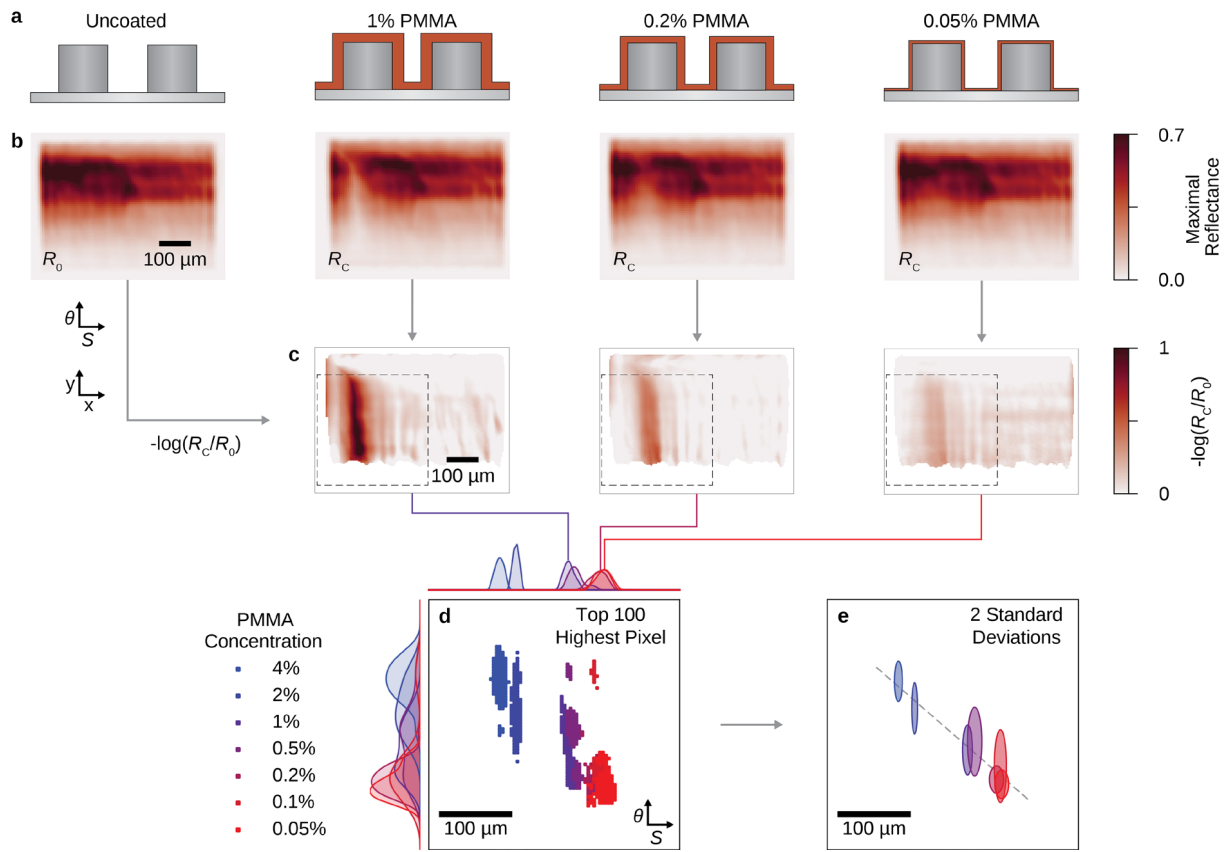

**Figure 5: Dual-gradient metasurfaces for molecular sensing.** (a) Unit cell sketches of the dual gradient with varying thicknesses of an analyte coating (PMMA). (b) Maximal reflectance maps of a dual gradient with $S = 0.95 - 1.1$ and $\theta = 0 - 45°$ for the different coating thicknesses following (a). From left to right, the maps show the gradient with no PMMA coating, followed by layers created with 1%, 0.2%, and 0.05% PMMA solutions. (c) The absorbance signal for each pixel is calculated using $-\log(R_C/R_0)$. The absorbance due to the analyte's vibrational fingerprint is evident within the left third of the dual gradient, with higher values corresponding to higher concentrations. (d) A zoomed-in section of the dual gradient, outlined by the black boxes in (c), presents the 100 pixels with the highest absorbance for seven different coating thicknesses. The range starts with thick layers (4%) represented in blue and ends with thin layers (0.05%) in red. The kernel density estimation of the pixel distributions is plotted along both the x- and y-axis. (e) Same zoomed-in section with the highest modulated pixels depicted as ellipses. The dimensions of the ellipses are set by twice the standard deviation in the x- and y-directions and are centered at the mean value. In gray the linear regression line.

# Discussion

In our study, we have investigated and established a fundamental methodology for describing the working principles of spectral BIC gradients and discussed their limitations in terms of scaling increments of neighboring unit cells. Our results on spectral gradients are of special importance in the context of miniaturizing optical systems with a possible key application of on-chip spectrometers. Here, the spectral gradient can be directly fabricated on top of, e.g. a CMOS sensor, allowing the metasurface to act as a two-dimensional equivalent of a grating or prism. This integration enables not only a simplified fabrication process with all components on a single chip but also a substantial reduction in system size.

We further introduce the concept of coupling gradients as a novel method to map the Q-factor of a resonance (and consequently its field enhancement) both continuously as well as spatially. We believe coupling gradients hold a high potential for studying emerging materials at the nanoscale due to its wide and continuous Q-factor tunability. These include controlling the density of photonic states in quantum systems within the framework of Fermi's golden rule and tuning the absorption of light in arbitrary target materials. For the latter, the balance between intrinsic losses within the material and radiative losses becomes crucial. The equilibrium point between these losses, known as critical coupling, is often the ideal state for applications as it allows for the highest field enhancement combined with maximal absorption, often reaching up to 100%. Our coupling-gradient method enables the continuous mapping of the entire coupling space, including the critical coupling point, without requiring prior detailed knowledge of the absorptive properties of the material under investigation. This approach could lead to significant breakthroughs in the discovery and fundamental understanding of new materials.

Finally, we have introduced the concept of dual gradients to achieve continuous coverage of the entire spectral and coupling parameter space. This advancement provides a straightforward platform to study a wide range of light-matter interactions. We particularly emphasize the potential impact in two key areas. First, the fundamental study of new materials, such as van der Waals materials or quantum dots, and their interaction with light. Numerous studies using BIC metasurfaces have demonstrated the enhancement of nonlinear properties, stimulated emission, and strong light-matter coupling in these materials[51,52,14]. Our dual-gradient metasurface can map these interactions continuously and within a compact area, which is crucial for research involving small footprint materials such as exfoliated 2D materials. The second research area concerns molecular sensing, where detection efficiency, but also miniaturization is of utmost importance. Our dual gradient achieves an unprecedented resonance density, and a total number of distinct modes of 27,500, all within a compact area of 650x450 µm². Our results, as shown in **Figure 4h,i**, further highlight how the sensitivity of a sensor is linked to the Q-factor and the concentration of the analyte. While high Q-factor resonances typically lead to increased sensitivity, they are also prone to losses within the resonator system. This can be problematic in scenarios with high analyte concentrations, which might saturate the resonance, and more critically, in the presence of other lossy, non-analyte materials, such as solvents. Particularly in medical applications, which often involve liquid environments, a trade-off between sensitivity and stability is necessary. Our dual gradient, with its complete coverage of the coupling space, consistently operates at optimal values, regardless of analyte concentration or the solvent.

In conclusion, our study marks a significant advancement in nanophotonics, offering practical solutions for miniaturizing optical systems such as on-chip spectrometers. The introduction of coupling gradients, along with the novel dual-gradient metasurface, enables precise control over light-matter interactions. This is particularly beneficial for the study of emerging materials like quantum dots and van der Waals materials, as well as for enhancing molecular sensing techniques. Our approach, with its ability to achieve high resonance density and sensitivity, holds great promise for applications in environmental monitoring, medical diagnostics, and potentially in the development of compact, efficient photocatalytic reactors.

## Materials and Methods

### Numerical methods

Our simulations were conducted using CST Studio Suite (Simulia), a commercial finite element solver. We configured the software for adaptive mesh refinement and periodic boundary conditions, operating in the frequency domain. The $CaF_2$ substrate was modeled as loss-free and non-dispersive within our targeted wavelength range with a refractive index of 1.36[53]. The refractive index of amorphous silicon was set to 3.32, based on near-infrared ellipsometric data.

### Sample fabrication

We deposited a 750 nm layer of amorphous silicon on a $CaF_2$ substrates using plasma-enhanced chemical vapor deposition (PECVD) with the PlasmaPro 100 system (Oxford Instruments). The nanostructuring process started with spin-coating a 400 nm layer of positive electron beam resist, ZEP520A (Zeon Corporation), followed by a conductive polymer coating using ESPACER (Showa Denko K.K). Electron beam lithography was performed using an eLINE Plus system (Raith) at 20 kV with a 20 μm aperture. The patterned films were developed in an amyl acetate bath, followed by a MIBK:IPA (1:9 ratio) bath. A 60 nm chromium layer was then deposited, and the resist was lifted off using Microposit Remover 1165 (Microresist). The remaining chromium served as an etching mask for the subsequent reactive ion etching process, which utilized SF6 and Argon gases. Finally, the chromium mask was removed using TechniEtch Cr01 (MicroChemicals).

### Optical characterization

Optical measurements were performed with a Spero spectral imaging MIR microscope (Daylight Solutions Inc.), as illustrated in **Figure S3a**. The microscope featured a 4x magnification objective ($NA = 0.15$) and provided a 2 mm² field of view with 480x480 pixel and a pixel size of approximately 4x4 μm². This system was equipped with three tunable quantum cascade lasers covering a wavelength range of 5.6-10.5 μm and offering a spectral resolution of 2 cm⁻¹. The lasers emitted linearly polarized light, essential for our measurements.

**Molecular Sensing**

We coated the dual gradient with different concentrations of PMMA (495 K, diluted in anisole), ranging from 0.05% to 4%. Each solution was uniformly applied to the metasurface through spin-coating at 3000 rpm for 1 minute. Following the coating, the metasurface was baked at 180 ºC for 3 minutes to ensure the PMMA layer was fully solidified. To dissolve the PMMA layer, the PMMA films were sequentially dissolved using an acetone bath, followed by an isopropyl alcohol (IPA) bath. As an additional cleaning step, the sample was UV cleaned for 30 minutes, followed by another round of acetone and IPA baths to remove any residual substances.

The thickness of the PMMA layers was estimated based on ellipsometric data obtained for 0.1%, 0.25%, and 4% solutions. We fitted a polynomial curve to this data with the following parameters: $d = 1.237 \cdot c^3 + 4.773 \cdot c^2 + 10.88 \cdot c + 0.9635$ with $d$ as the thickness in nm and $c$ as the concentration in %. This leads to layer thicknesses ranging from 1.5 nm (0.05%) to 200 nm (4%). The measurements were conducted using an HS-190 J.A. Woollam VASE ellipsometer.

# Acknowledgments

This project was funded by the Deutsche Forschungsgemeinschaft (DFG, German Research Foundation) under grant numbers EXC 2089/1–390776260 (Germany's Excellence Strategy) and TI 1063/1 (Emmy Noether Program), the Bavarian program Solar Energies Go Hybrid (SolTech) and the Center for NanoScience (CeNS). Funded by the European Union (METANEXT, 101078018 and NEHO, 101046329). Views and opinions expressed are however those of the author(s) only and do not necessarily reflect those of the European Union or the European Research Council Executive Agency. Neither the European Union nor the granting authority can be held responsible for them. S.A.M. additionally acknowledges the Lee-Lucas Chair in Physics and the EPSRC (EP/W017075/1).

# Author contributions

A.A. and A.T conceived the idea and planned the research. A.A and T.W contributed to the sample fabrication. A.A. performed the measurements. A.A. conducted the numerical simulations. A.A., T.W., and A.W. contributed in the data processing. A.A., T.W., S.A.M., and A.T. contributed to the data analysis. S.A.M. and A.T. supervised the project. All authors contributed to the writing of the paper.

# Competing interests

Authors declare that they have no competing interests.

# Data and materials availability

All data are available in the main text or the supplementary materials.

# Supplementary Information:

# Continuous spectral and coupling-strength encoding with dual-gradient metasurfaces

Andreas Aigner[1], Thomas Weber[1], Alwin Wester[1], Stefan A. Maier[1,2], and Andreas Tittl[1,*]

[1]Chair in Hybrid Nanosystems, Nano-Institute Munich, Faculty of Physics, Ludwig-Maximilians-Universtität München, Königinstraße 10, 80539 München, Germany

[2]School of Physics and Astronomy, Monash University, Wellington Rd, Clayton VIC 3800, Australia

[3]The Blackett Laboratory, Department of Physics, Imperial College London, London, SW7 2AZ, United Kingdom

*Corresponding author. Email: Andreas.Tittl@physik.uni-muenchen.de

## Table of contents

# Supplementary Notes

**Note 1: Temporal coupled mode theory and fitting**

Temporal coupled mode theory (TCMT) is a straightforward and well-established formalism for describing coupled resonances in cavities that are also connected to the far field[1,2]. In our study, we employ a model consisting of a single resonator coupled to two ports, enabling transmission and reflection (**Figure S6**). The excitation occurs through port 1, represented as $\boldsymbol{s}_+ = (s_{1+}, 0)^T$ and the output can be described as $\boldsymbol{s}_- = (s_{1-}, s_{2-})^T$, where $s_{1-}$ denotes the reflected and $s_{2-}$ the transmitted waves. The time evolution of the resonant mode amplitude $a(t)$ of our mirror-symmetric metasurface is described by the following coupled equations:

$$\frac{da(t)}{dt} = (i\omega_0 - \gamma_{tot})\, a(t) + \boldsymbol{\kappa}^T \boldsymbol{s}_+ \quad (S1)$$

$$\boldsymbol{s}_- = C\boldsymbol{s}_+ + a(t)\boldsymbol{\kappa} \quad (S2)$$

Here, the resonance frequency of the mode is denoted by $\omega_0$, the total damping rate consists of the radiative damping rate and parasitic damping rates such as material absorption and scattering losses $\gamma_{tot} = \gamma_{rad} + \gamma_{int} + \gamma_{scat} = \gamma_{rad} + \gamma_{para}$. The coupling between the ports and the mode is governed by the radiative damping rate $\boldsymbol{\kappa} = \left(\sqrt{\gamma_{rad}}, \sqrt{\gamma_{rad}}\right)^T$, whereas the non-resonant coupling between the port is described by the unitary matrix $C$

$$C = e^{i\varphi} \cdot \begin{pmatrix} r_0 & it_0 \\ it_0 & r_0 \end{pmatrix} \quad (S3)$$

where $r_0$ ($t_0$) describes the background reflection (transmission) with $r_0^2 + t_0^2 = 1$ and $\varphi$ is a global phase, together giving rise to asymmetric Fano lineshapes.

Assuming a time-harmonic mode amplitude ($a(t) \sim e^{i\omega t} \rightarrow \dot{a} = i\omega a$), we can rewrite equations (S1) and (S2) and solve for the reflection coefficient

$$r(\omega) = \frac{s_{1-}}{s_{1+}} = e^{i\varphi} r_0 + \frac{\gamma_{rad}}{i(\omega - \omega_0) + \gamma_{tot}}$$

The reflectance signals can thus be fitted by $R(\omega) = |r(\omega)|^2$ and the total Q-factor can be calculated from $Q_{tot} = \frac{\omega_0}{2\gamma_{tot}}$.

Next, we calculate the mode amplitude $a$ as

$$a(\omega) = \frac{\sqrt{\gamma_{rad}}\, s_{1+}}{i(\omega - \omega_0) + \gamma_{para} + \gamma_{rad}}$$

Because $a$ is normalized such that $|a|^2$ denotes the total energy in the cavity, the electric field enhancement can be expressed as

$$\left|\frac{E}{E_0}\right|^2 \sim \left|\frac{a_{max}}{s_{1+}}\right|^2 = \frac{\gamma_{rad}^2}{\left(\gamma_{para} + \gamma_{rad}\right)^2}$$

Where $a_{max} = a(\omega = \omega_0)$ is the maximum amplitude at resonance frequency.

This is the general expression for a metasurface exhibiting both radiative and parasitic losses.

For lossless systems or at the critical coupling condition ($\gamma_{para} = \gamma_{rad}$), the dependence of the field enhancement simplifies to

$$\left|\frac{E}{E_0}\right|^2 \sim \frac{1}{\gamma} \sim Q$$

Where $Q$ is either the radiative Q-factor (lossless case) or total Q-factor at the critical coupling condition. Thus, the Q-factor of our metasurface has direct and linear influence on the electric near-field enhancement.

**Note 2: Effect of the unit cell arrangement in a scaling gradient**

At its core, the spectral gradient can be interpreted as a concatenation of individual resonant chains, all with slightly offset resonance wavelengths. A single chain of resonators has been shown to produce a weak but observable BIC resonance when the resonators are positioned such that their effective dipole moments align, allowing them to form a collective dipole[3,4]. Therefore, analyzing and understanding the impact on the unit cells' orientation in these chains within the gradient is crucial for the optical performance.

To investigate this effect in our spectral-gradient metasurfaces, we fabricated them in two configurations: one with scaling perpendicular to the excitation polarization (**Figure S5a**), as shown in **Figure 2**, and another with scaling parallel to the excitation polarization (**Figure S5b**). In our tilted double-ellipse design, the resonance emerges due to a net dipole moment along the unit cells' long axis indicated by the green arrow in **Figure S5a**. This is generated by two opposing individual dipoles (black arrows) within the ellipses when the tilting angle $\theta$ is non-zero. Consequently, a collective excitation occurs when the scaling is perpendicular to the excitation polarization. However, when the scaling is parallel to the excitation polarization, as depicted in **Figure S5b**, a collective dipole cannot form easily. This is due to the y-alignment offset of unit cells and the spectral shift of the dipolar moments along the coupling direction, which results in poor spectral overlap.

We experimentally tested both configurations and further examined the influence of gradient steepness, essentially the total spectral range within a spectral gradient of a given dimension. For this, we fabricated spectral gradients measuring 600x150 μm$^2$ with varied spectral ranges, spanning 225 nm to 1910 nm. **Figure S5c** (equivalent to **Figure 2g**) and **Figure S5d** display the average maximal reflectance amplitude for the gradients of different steepness in the two configurations. Strikingly, gradients with scaling normal to the excitation polarization exhibit a significantly better performance, for some spectral ranges more than twice the amplitude. This confirms the importance of the proper alignment of unit cells, ensuring effective dipole moment coupling, for optimal gradient performance.

Furthermore, the $\varepsilon_s$-dependency of amplitude and Q-factor for both configurations is plotted in **Figure S5e,f**. Solid lines represent gradients with scaling normal to the excitation polarization revealing their superior performance, while the dashed red lines show the results for metasurfaces with scaling parallel to the excitation polarization.

**Note 3: Influence of scattering loss on the position of highest field enhancement**

As discussed in **Supplementary Note 1**, in lossless systems, or at the critical coupling position the field enhancement is proportional to the Q-factor via $FE^2 \sim Q$. This implies the highest field enhancement coincides with the maximal Q-factor. However, this relationship becomes more complex when considering intrinsic and scattering losses, as further detailed in **Supplementary Note 1**.

In scenarios involving losses, the point of maximal field enhancement shifts to a lower Q-factor, known as the critical coupling point where $\gamma_{\text{int}} = \gamma_{\text{rad}}$. At this point, both resonance amplitude and absorption are maximized. While this phenomenon is extensively studied in lossy systems like plasmonic metasurfaces, it is less frequently addressed in dielectric systems, where intrinsic losses are typically minimal. Nevertheless, dielectric metasurfaces encounter additional non-intrinsic loss channels, collectively termed as scattering losses $\gamma_{\text{scat}}$ in **Supplementary Note 1**. These losses arise from various sources such as resonator size variations, surface roughness, material inhomogeneities, finite array size, limited illumination area, and imperfect light collimation.

The point of highest resonance amplitude and thus field enhancement within our gradient is determined by specific fabrication qualities and the measurement setup. To accurately identify this point, we analyzed spectra along the gradient, dividing the gradient into 31 subsections, each representing a $\theta$ range of approximately 1.45° of the total 45°. The outermost sections were excluded due to scattering losses caused by the proximity to the edges. Given that structural quality, surface roughness, intrinsic losses, and losses from non-collimated excitations remain consistent across the gradient, we treated the parasitic losses, the combined intrinsic and scattering losses $\gamma_{\text{para}} = \gamma_{\text{int}} + \gamma_{\text{scat}}$, as constant. **Figure S9a** displays all spectra with their fits, demonstrating a good agreement across all $\theta$ values. Note that this fitting procedure with a shared parameter $\gamma_{\text{para}}$ is computationally way more demanding than single spectra fits, as it is based on iteratively fitting all spectra in order to minimize the error for $\gamma_{\text{para}}$. **Figure S9b** illustrates the $\theta$-dependent parasitic and radiative losses, and **Figure S9c** presents the corresponding Q-factors. The critical point of our analysis is the crossing of radiative and parasitic losses as well as radiative and parasitic Q-factors, indicating a critically coupled system with maximal field enhancement and highest coupling to adjacent materials. In our spectrally-aligned coupling gradient, this critical point is approximately at $\theta = 9°$.

This 9° deviation from the ideal scenario, where the highest field enhancement is near 0°, we attribute to our fabrication techniques and measurement setup. While intrinsic losses of silicon in the mid-infrared range are negligible, we assume that the primary source of parasitic losses stems from our measurement setup, particularly the non-perfectly collimated light.

It is important to clarify that these limitations are not fundamental and can be overcome in future applications of coupling-gradient metasurfaces. The exact position of this critical point may vary depending on intrinsic losses of the resonator material or any material of interest in proximity. For our current results, this implies that the effective coupling gradient begins not at the very end of the gradient ($\theta = 0°$) but slightly inward, around 9°. Beyond this point, for the remaining 36° $\theta$-sweep, we observe a continuous, smooth coupling sweep, as visible in **Figure S9b**.

Finally, it is crucial to recognize that the point of highest field enhancement is not universally the most desirable. In applications like molecular sensing, high sensitivity does not necessarily correlate with the point of strongest field enhancement. While a robust field enhancement suggests strong coupling between the analyte and the resonator, a resonance that is sensitive to analyte-induced losses is often more critical. Hence, high sensitivity sensors tend to favor higher Q-factors over maximum field enhancement.

**Note 4: Calculating the resonance density**

In the spectral gradient, both $P_x$ and $P_y$ are scaled by the same factor $S$ along the x-direction. The number of columns along the y-direction will vary depending on the scaling factor at each point along the x-axis. The number of unit cells $N_{\text{unit cell y}}$ within a column of the metasurface with height $L_y$ for a specific scaling factor $S$ can be calculated via

$$N_{\text{unit cell y}}(S) = \left\lfloor \frac{L_y}{S \cdot P_y} \right\rfloor$$

with "$\lfloor\ \rfloor$" denoting the floor function, which rounds off to the nearest integer, ensuring that only whole unit cells within the border of the metasurface are counted.

Since the scaling changes linearly along the x-axis, one can calculate the number of unit cells along the x-axis by considering the average scaling factor. Let $S_{\text{start}}$ and $S_{\text{end}}$ be the start and end values of the scaling factor, respectively. The average scaling factor $S_{\text{avg}}$ is given by $S_{\text{avg}} = \frac{1}{2} \cdot (S_{\text{start}} + S_{\text{end}})$.

Then, the number of unit cells along the x-axis is calculated by dividing the total length of the metasurface in x-direction ($L_x$) by the pitch in the x-direction scaled by the average scaling factor resulting in

$$N_{\text{unit cell x}} = \left\lfloor \frac{L_x}{S_{\text{avg}} \cdot P_x} \right\rfloor.$$

The total number of unit cells is obtained by integrating $N_{\text{unit cell y}}(S)$ over the range of scaling factors:

$$N_{\text{unit cell}} = \int_{S_{\text{start}}}^{S_{\text{end}}} N_{\text{unit cell y}}(S)\, dS$$

Given the linear variation of $S$, this integral simplifies to

$$N_{\text{unit cell}} = N_{\text{unit cell}} \cdot N_{\text{unit cell y,avg}}.$$

With $N_{y,\text{avg}}$ as the estimated average number of unit cells along the y-axis given by $N_{y,\text{avg}} = \frac{1}{2} \cdot (N_{\text{unit cell y,start}} + N_{\text{unit cell y,end}}) = \left\lfloor \frac{1}{2} \cdot \left( \frac{L_y}{S_{\text{start}} \cdot P_y} + \frac{L_y}{S_{\text{end}} \cdot P_y} \right) \right\rfloor$ which finally leads to

$$N_{\text{unit cell}} = N_{\text{unit cell}} \left\lfloor \frac{1}{2} \cdot \left( \frac{L_y}{S_{\text{start}} \cdot P_y} + \frac{L_y}{S_{\text{end}} \cdot P_y} \right) \right\rfloor.$$

For the investigated spectral gradient in **Figure 2** with $P_x = 2.4$ µm, $P_y = 4$ µm, $L_x = 600$ µm, $L_y = 150$ µm, $S_{\text{start}} = 1$, and $S_{\text{end}} = 1.1$, the number of unit cells along the x- and y-dimensions compute to $N_x = 238$, $N_{\text{unit cell y,avg}} = 35$, while the total number of unit cells is $N_{\text{unit cell}} = 8330$.

As discussed in the main text for spectral gradients in **Figure 2** the resonance density per unit cell can be calculated via

$$\rho_{res} = \frac{N_{unit\ cell\ x}}{N_{unit\ cell}} = \frac{1}{N_{unit\ cell}} = 2.9 \cdot 10^{-2}.$$

The coupling gradient presented in **Figure 3** has the dimensions $P_x = 2.4$ µm, $P_y = 4$ µm, $L_x = 650$ µm, $L_y = 150$ µm, the same number of unit cells $N_{unit\ cell\ y}$ in each column, and $\Delta S = 0$, so $S_{start} = S_{end} = 1$. $\rho_{res}$ can be calculated, analog to the scaling gradient, via

$$\rho_{res} = \frac{N_{unit\ cell\ x}}{N_{unit\ cell}} = \frac{1}{N_{unit\ cell\ y}} = 2.7 \cdot 10^{-2}.$$

For the dual gradient discussed in **Figure 4a-d**, each unit cell slightly differs from all other unit cells, resulting in an equal number of unit cells $N_{total}$ and resonances within the gradient, thus

$$\rho_{res} = 1.$$

In comparison, the monospectral gradient with $S = 1.05$ from **Figure 2** gives $\rho_{res} = 4.7 \cdot 10^{-4}$. This means an improvement of the resonance density per unit cell $\rho_{res}$ by a factor of 62, 57, and 2128 for the spectral, the coupling, and the dual gradient, respectively.

**Note 5: Resonance density comparison to literature**

In our literature comparison, we focused on extended structures like metasurfaces and gratings to benchmark the resonance density of our gradient metasurfaces, as individual resonators are challenging to compare. Although single resonators can exhibit a high resonance density, when arranged into an array they need large interparticle distances to avoid scattering effects. This significantly lowers the overall density of resonators per unit area. The ideal interparticle distance varies and lacks a definitive standard, complicating the establishment of a reliable resonance density metric that accounts for this factor. However, our subwavelength arrangement, where each resonator has a unique resonance, suggests that our dual-gradient metasurfaces significantly surpass the resonance density of arrays composed of individual, non-coupling resonators.

In **Table 1**, we present key parameters of the resonant nanostructures under consideration, including their Q-factor, estimated number of resonances, and resonance density. Calculating these values is challenging due to the variety of the systems analyzed. We attempted to estimate these parameters as accurately as possible, which involved treating 1D gratings as 2D metasurfaces with quadratic unit cells. This approach allowed us to calculate the effective total number of unit cells within the structure, providing a basis for our comparisons.

# Supplementary Tables

**Table 1: Comparison of the total number of resonances and the resonance density.**

| Resonator system | Resonance type | Spectral range | Exp. Q-factor | Number of resonances | Resonance density | Reference |
|---|---|---|---|---|---|---|
| **Individual spectral gradient (our work)** | BIC | Mid-IR | 80 | **238** | $2.9 \cdot 10^{-2}$ | - |
| **Individual coupling gradient (our work)** | BIC | Mid-IR | 170 | **270** | $2.7 \cdot 10^{-2}$ | - |
| **Dual gradient (our work)** | BIC | Mid-IR | 170 | **27,500** | **1** | - |
| Molecular sensor for SEIRA | BIC | Mid-IR | 120 | 100 | $9.1 \cdot 10^{-4}$ | [5] |
| Pixelated RI sensor | BIC | Vis | 180 | 12 | $2.9 \cdot 10^{-6}$ | [6] |
| Trapped Rainbow | Plasmonic grating | Vis | 30 | 116 | $1.9 \cdot 10^{-2}$ | [7] |
| Multiple Surface lattice resonances | Plasmonic SLR | Near-IR | - | - | $1.7 \cdot 10^{-4}$ | [8] |
| Plasmonic gradient | Plasmonic resonators | Vis | 10 | - | $2.5 \cdot 10^{-2}$ | [9] |
| Dielectric spectral-gradient metasurface | BIC | Near-IR | 75 | 1,476 | $4.3 \cdot 10^{-3}$ | [10] |
| Radial bound states in the continuum | BIC | Vis | 500 | - | $3.5 \cdot 10^{-2}$ | [3] |

# Supplementary Videos

All videos were directly taken from the hyperspectral measurements, after a rotation by 20°, and cropping the displayed field of few.

**Supplementary Video 1: Spectral gradient**

Reflectance video of the three monospectral metasurfaces with scaling factors of 1.01, 1.05, and 1.09, and the spectral gradient of $S = 1.0 - 1.1$, shown in **Figure 2**. The reflectance color code spans values from 0 to 0.7, the start frame is at 5672 nm, and the end frame at 6569 nm.

**Supplementary Video 2: Coupling gradient**

Reflectance video of the coupling gradient (top) and the spectrally aligned coupling gradient (bottom), discussed in **Figure 3**. The reflectance color code spans values from 0 to 0.9, the start frame is at 5608 nm, and the end frame at 6319 nm.

**Supplementary Video 3: Dual gradient**

Reflectance video of the dual gradient discussed in **Figure 4**. The reflectance color code spans values from 0 to 0.7, the start frame is at 5574 nm, and the end frame at 7153 nm.

**Supplementary Video 4: Dual gradient for molecular sensing, uncoated**

Reflectance video of the uncoated dual gradient discussed in **Figure 5**. The reflectance color code spans values from 0 to 0.7, the start frame is at 5574 nm, and the end frame at 6588 nm.

**Supplementary Video 5: Dual gradient for molecular sensing, coated**

Reflectance video of the dual gradient discussed in **Figure 5,** coated with a 1% solution of PMMA. The reflectance color code spans values from 0 to 0.7, the start frame is at 5574 nm, and the end frame at 6588 nm.

# Supplementary Figures

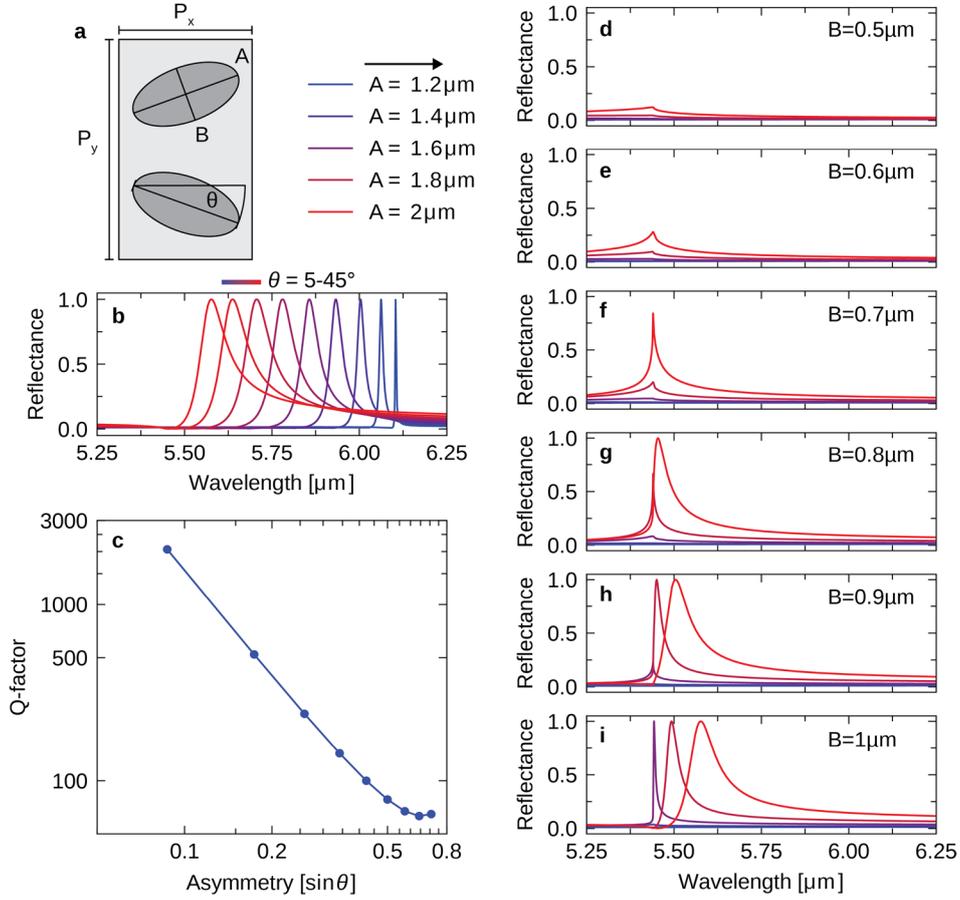

**Figure S1: Numerical analysis of the dual-ellipse geometry.** (a) Illustration of the tilted dual-ellipse unit cell, showing the short pitch $P_x$ along the x-axis, and the long pitch $P_y$ along the y-axis. The ellipse diameters of the long and short axes are labeled as $A$ and $B$, respectively, with $\theta$ representing the tilt angle of the ellipse's long axis $A$ relative to the x-axis. (b) Display of the tilt angle sweep, starting from $\theta = 5°$ (blue) to $\theta = 45°$ (red) in 5° increments. (c) Q-factors derived from (b) using TCMT fitting. Both axes are logarithmic, illustrating the typical symmetry-protected BIC relationship of $Q = \frac{1}{\alpha^2} = \frac{1}{sin^2(\theta)}$ as a linear plot with a slope of approximately -2. Deviations from this linear trend at large angles are attributed to the influence of the Rayleigh anomaly, as the BIC shifts towards it with increasing $\alpha$. To achieve a wide range of Q-factors, we simulated 45° tilted ellipses of varying diameters $A$ and $B$, while maintaining constant values for $P_x$, $P_y$, and the height. (d) – (i) Reflectance spectra for varying $A$ from 1.2 to 2 µm in 5 steps and $B$ from 0.5 to 1 µm in 6 steps. It's evident that combinations of small $A$ and $B$ do not exhibit a resonance, as they are blueshifted towards the Rayleigh anomaly, allowing energy to radiate away via the first diffraction order, preventing the formation of a strong BIC mode. Visible BIC resonances are only observed for larger $A$ and $B$ values. However, most visible modes have relatively high Q-factors due to their proximity to the Rayleigh anomaly[11]. The resonance for $A = 2$ µm and $B = 1$ µm is the only one with a sufficiently broad lineshape for our purposes. Larger structures were not considered to avoid

overlap at the tips of the ellipses at $\theta = 45°$. As shown in (c), the Q-factor deviation for high angles does not significantly stray from the ideal Q-factor – asymmetry relation.

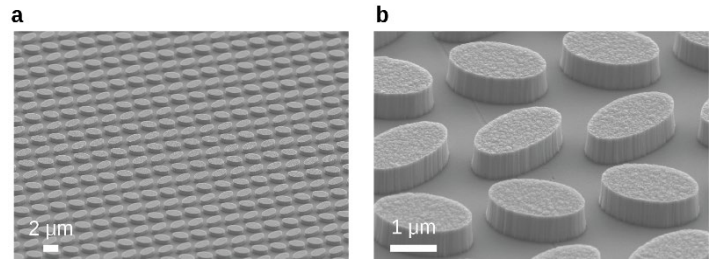

**Figure S2: Spectral gradient SEM images.** SEM images of the spectral gradient investigated in Figure 2 with $\Delta S = 0.1$, acquired at a tilting angle of 45°. Overview image in (a) and close-up in (b).

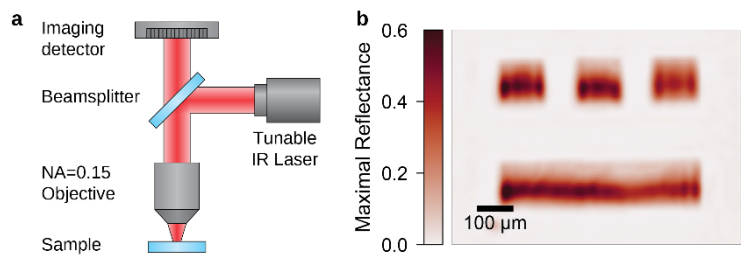

**Figure S3: Optical characterization and reflectance map.** (a) Sketch of the measurement setup. The light of three tunable quantum cascade lasers is guided via a dichroic mirror onto the sample through a 4x, $NA = 0.15$ objective. The reflected light is collected by the same objective and projected onto a 480x480 pixel detector. The lasers are adjusted in 2 cm$^{-1}$ steps across the target spectral range, and an image is captured at each wavelength, resulting in a hyperspectral image. (b) Display of the maximum reflectance of each pixel of the spectral gradient ($\Delta S = 0.1$) across wavelengths from 5.6 to 7 μm, illustrating consistently high reflectance amplitudes across the gradient.

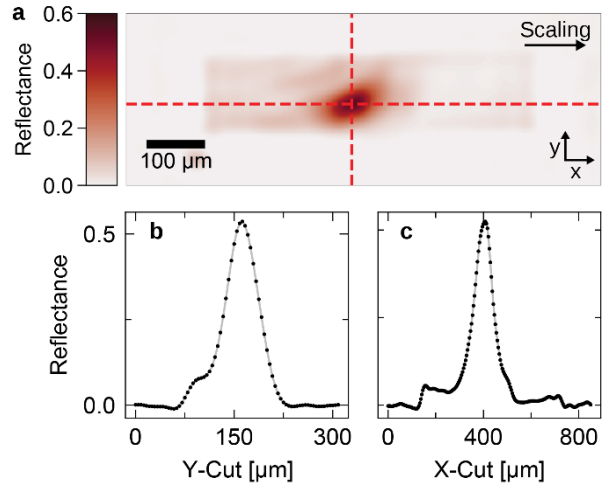

**Figure S4: Cuts of the reflectance map of the spectral gradient.** (a) Reflectance snapshot at 6.1 μm for the spectral gradient depicted in Figure 2 with $\Delta S = 0.1$. Specific cuts from the reflectance map in (a), corresponding to the two red dashed lines. (b) Vertical section along the y-axis, and (c) horizontal section along the x-axis.

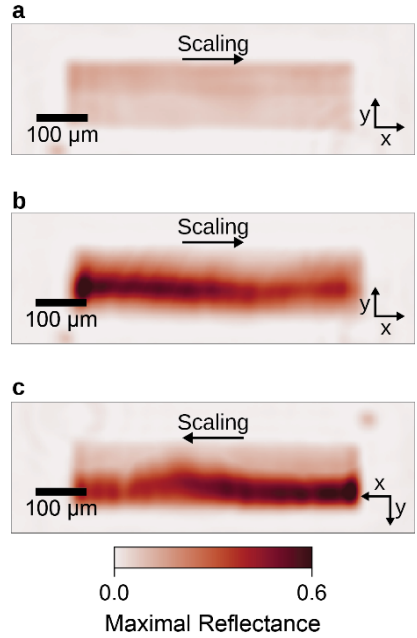

**Figure S5: Effects of rotated excitation polarization on a spectral gradient.** Maximal reflectance maps of a spectral gradient with $\Delta S = 0.1$. In panel (a), the polarization of the electric field is rotated by 90° and is now parallel to the x-axis. Under this condition, no significant reflectance revealing a resonance is observed. The minor increase in reflectance within the gradient is attributed to the structured Si layer rather than the resonant effect. Panels (b) and (c) show the electric field polarization along the y-axis, analogous to the configuration in Figure 2, where high reflectance values are observed due to the excited BIC resonance. In panel (c), the gradient is rotated by 180°. Notably, regions on the lower half of the resonator exhibit higher reflectance values in both orientations. This observation suggests that the lower reflectance observed on the upper half is not a consequence of the spectral gradient itself, but rather a side effect of the experimental setup.

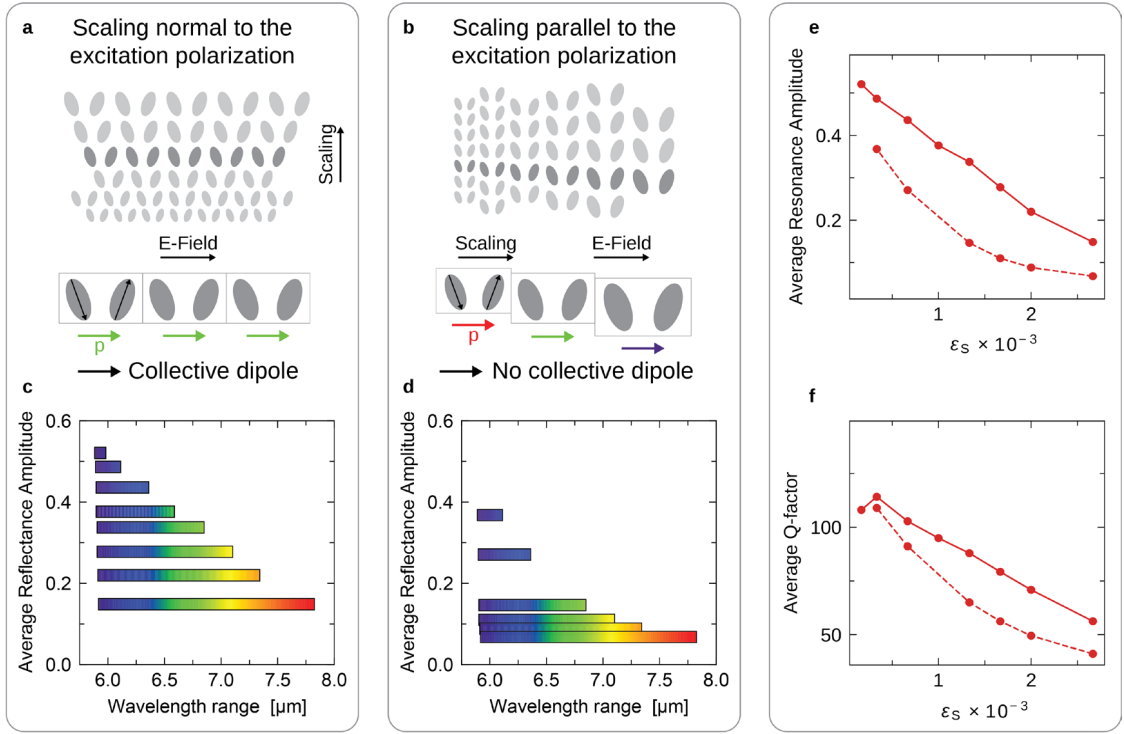

**Figure S6: Scaling normal and parallel to the excitation direction.** (a) Illustration of a spectral gradient with scaling perpendicular to the excitation polarization, allowing the formation of a collective dipolar mode across a chain of identical resonators. (b) Illustration of a spectral gradient with scaling parallel to the excitation polarization. This configuration hinders the collective dipole formation due to y-position mismatches in adjacent unit cells and their resonance wavelength offsets. (c) Graph showcasing the average reflectance amplitude and wavelength range for gradients with perpendicular scaling as in (a). Peak reflectance values correspond to the shortest wavelength range. (d) Average reflectance amplitude, analog to (c), but for gradients with parallel scaling as in (b). This configuration exhibits a notably reduced performance compared to the perpendicular scaling in (c). (e) Average resonance amplitude plotted against different $\varepsilon_s$ for gradients with $\theta = 20°$. Solid lines represent perpendicular scaling to the excitation polarization, while dashed lines indicate parallel scaling. (f) Average Q-factors derived from TCMT fits for the spectral gradients, analog to (e).

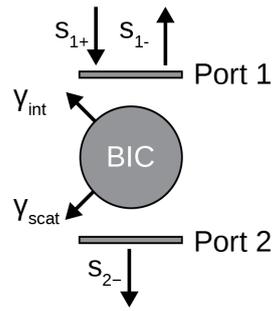

**Figure S7: TCMT scheme.** Schematic illustration of the used TCMT model consisting of a cavity which supports one resonance (BIC), two ports, one representing transmitted and one representing reflected light. Additionally, the cavity is connected to two parasitic loss channels, loss channels, the intrinsic loss $\gamma_{\text{int}}$, and scattering loss $\gamma_{\text{scat}}$.

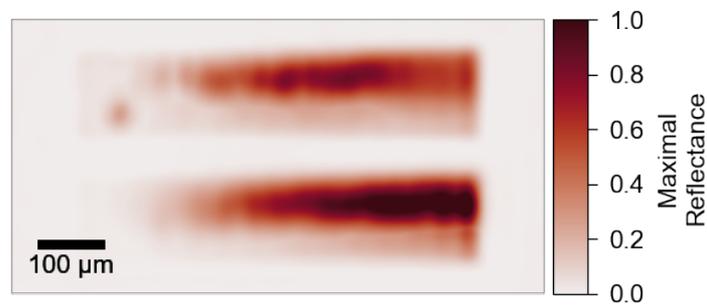

**Figure S8: Maximal reflectance map of coupling gradient.** Maximal reflectance value of each pixel of the two coupling gradients, presented in Figure 3. On the top the normal coupling gradient, and at the bottom, the spectrally-aligned coupling gradient.

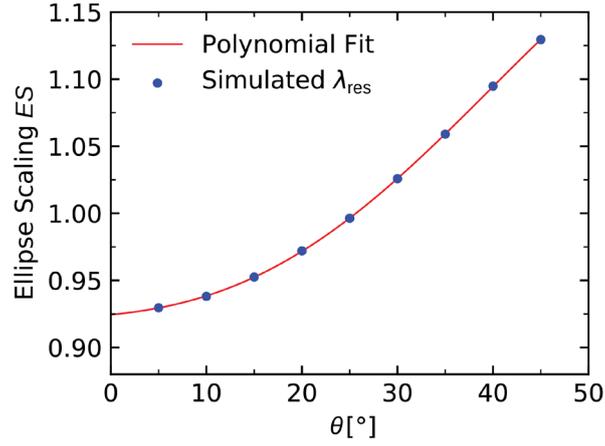

**Figure S9: Dispersion correction for $\theta$.** Numerically retrieved ellipse scaling factors *ES* for periodic metasurfaces of varying ellipse tilting angle $\theta$ from 5 to 45° pictured as blue dots. The scaling factors were retrieved by finely sweeping *ES* for all $\theta$ to achieve resonances precisely at 5.8 µm. The numerical results were then fitted by a second order polynomial function, pictured as a red line, yielding: $y = 8.24 \cdot 10^{-5} \cdot x^2 + 9.99 \cdot 10^{-4} + 0.921$. These results were then used to fabricate the spectrally aligned coupling gradient in Figure 3f-k and the dual gradients in Figure 4.

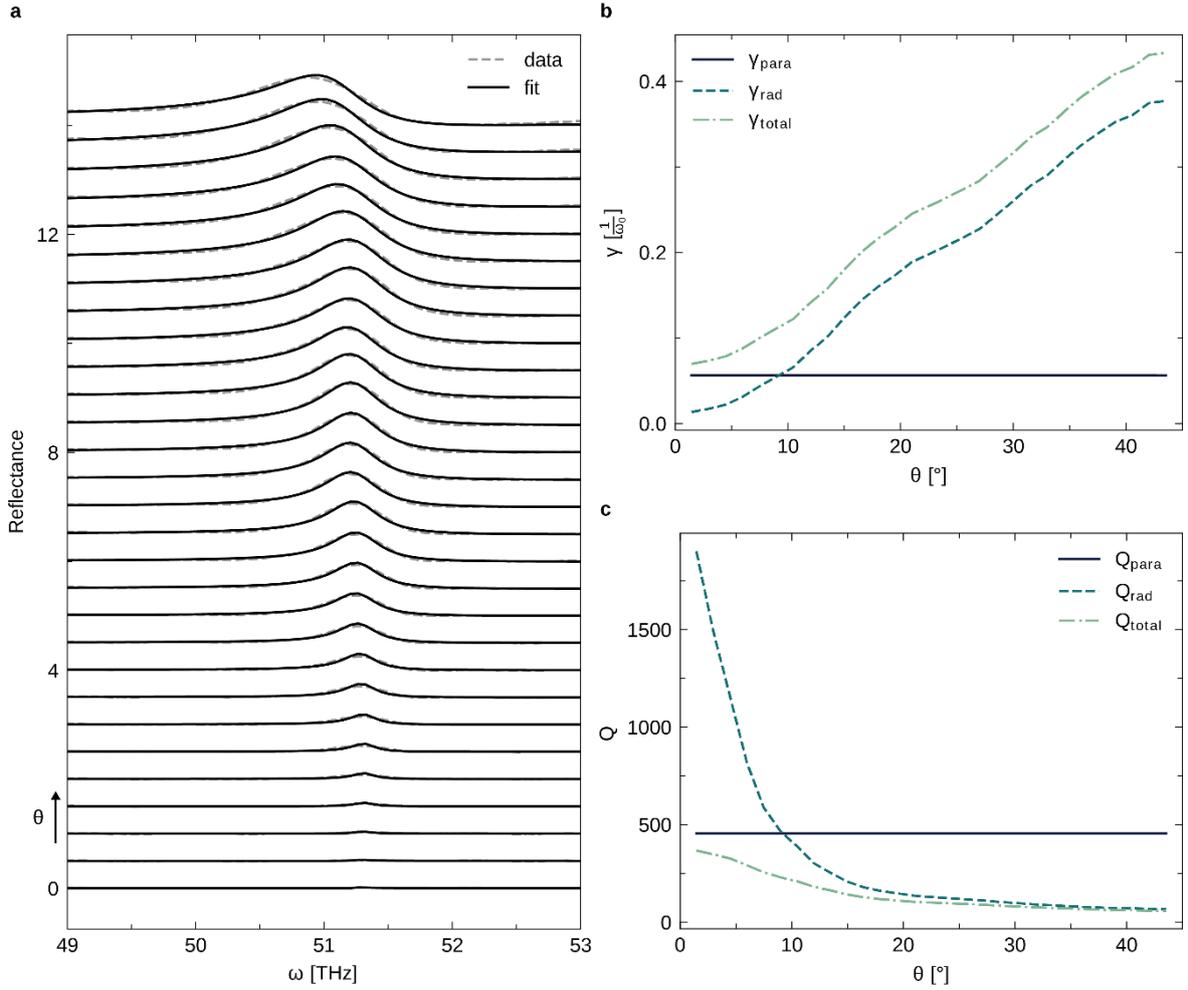

**Figure S10: Study on parasitic losses in the coupling gradient.** (a) Reflectance spectra of the spectrally-aligned coupling gradient of Figure 3f-k as gray dashed lines and the corresponding TCMT fits as solid black lines for a fixed $\gamma_{para}$. The spectra are taken along the x-axis of the gradient, with in total 29 steps (one step on each side of the gradient is excluded due to etch effects), each step roughly corresponding to a change of $\theta$ by 1.45°, the spectra are offset by 0.5 for clarity. (b) Extracted loss rates $\gamma_{para}$ and $\gamma_{rad}$, and the combined loss rate $\gamma_{total}$ plotted against the corresponding $\theta$ values. A clear crossing for $\gamma_{para}$ and $\gamma_{rad}$ is visible around $\theta = 9°$. (c) Extracted Q-factors $Q_{para}$, $Q_{rad}$, and $Q_{total}$, analog to (b).

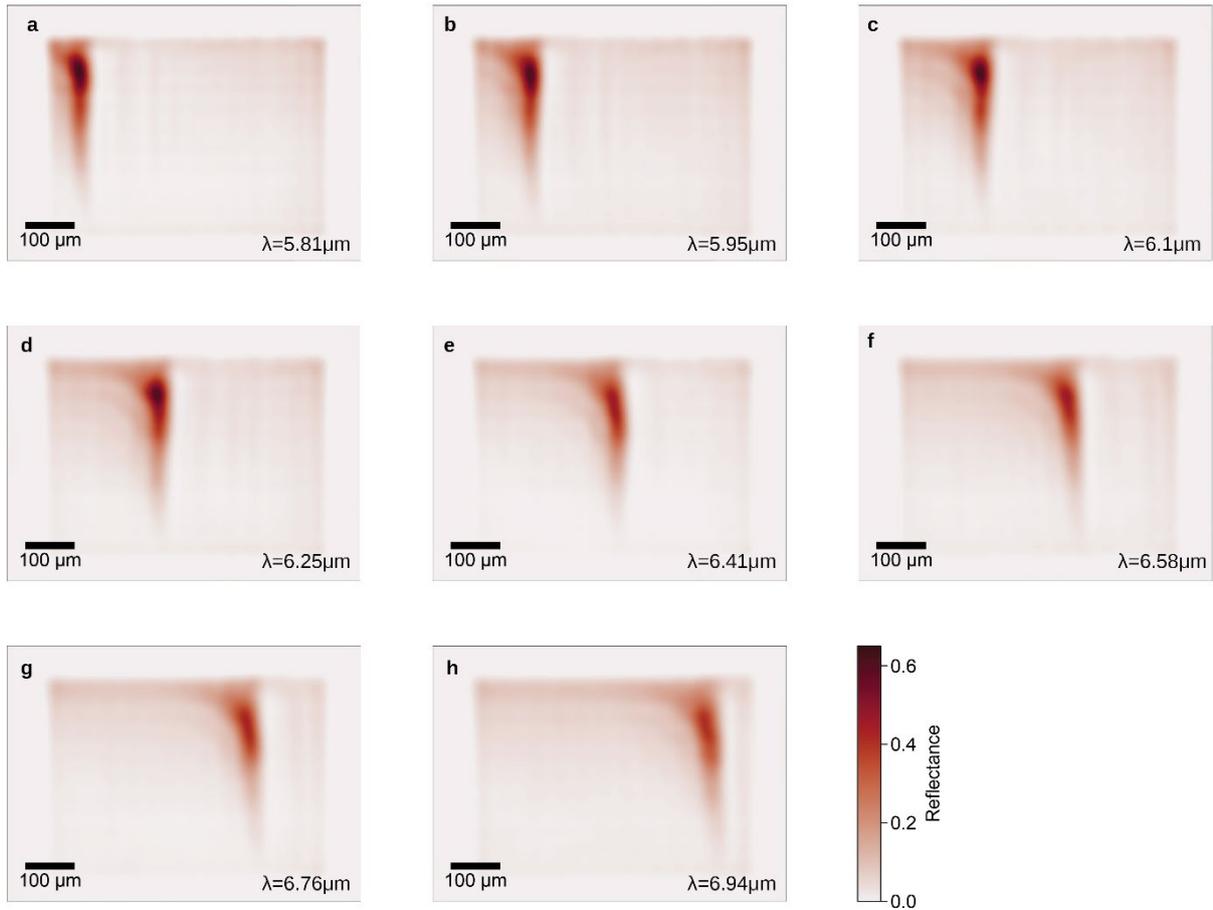

**Figure S11: Single wavelength snapshots of the dual-gradient metasurface.** Single wavelength snapshots of reflected light of the dual gradient discussed in Figure 4a-d with $S = 0.95 - 1.25$ and $\theta = 0 - 45°$ at a wavelength ranging from 5.81 in (a) to 6.94 μm in (h). Resonant sections within the gradient are apparent from high reflectance values.

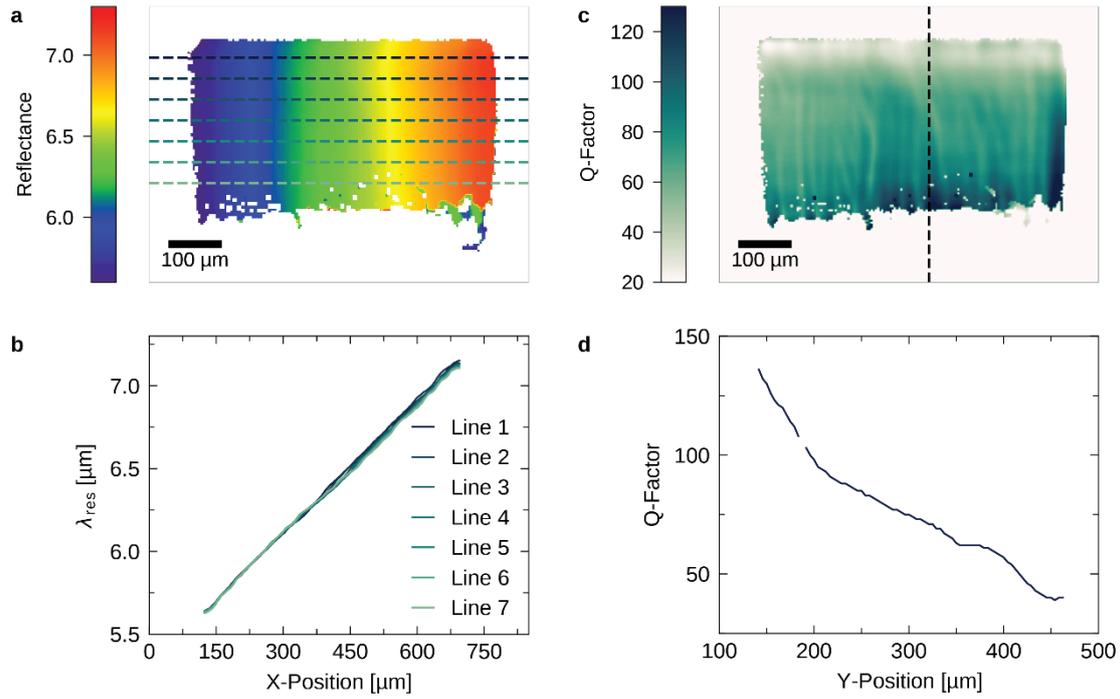

**Figure S12: Resonance frequency and Q-factor slices of the $\Delta S = 0.3$ dual gradient.** (a) Resonance frequency map of the dual gradient shown in Figure 4a-d. The seven horizontal dashed lines with a spacing of 10 pixels indicate the cuts taken along the scaling direction of the dual gradient. (b) Cuts of the map shown in (a) with the resonance wavelength plotted against the x-position. All seven plots show linear and uniform resonance wavelength shifts across the gradient, independent of the y-position within the gradient. (c) Q-factor map of the dual gradient shown in Figure 4a-d. (d) Q-factor cut along the y-axis of the map shown in (c) where the cut is indicated as a dashed line. A clear trend towards lower Q-factors for increasing asymmetry (increasing y-position) is visible. One point was removed from the cut, due to a fitting irregularity ($Q > 2000$).

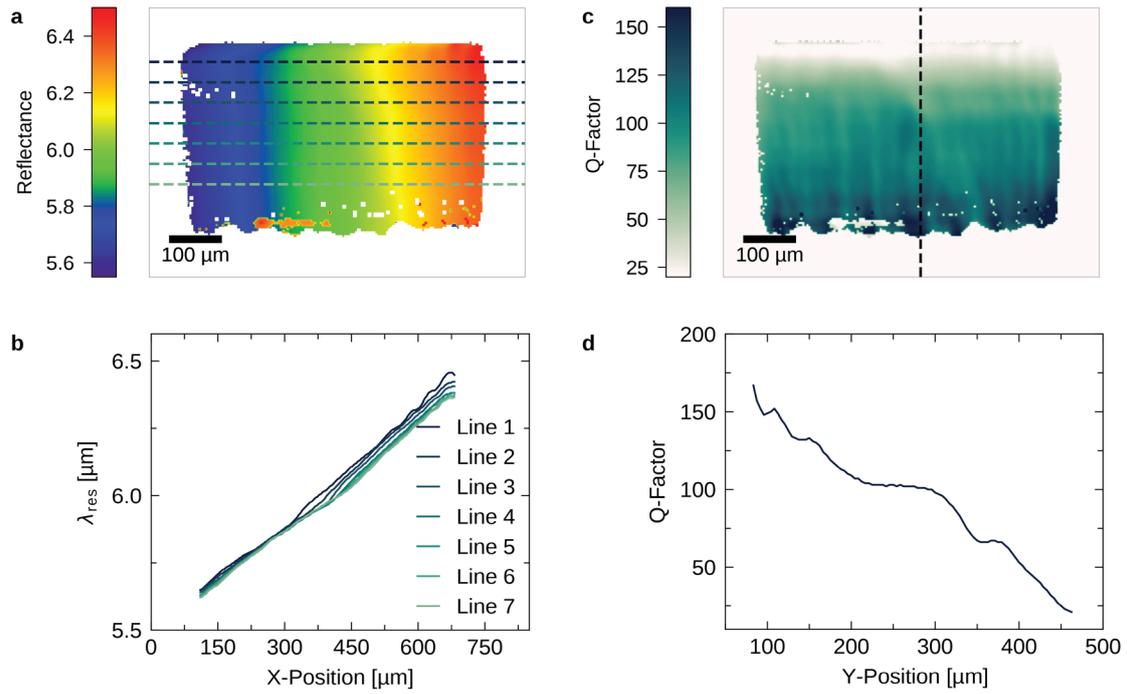

**Figure S13: Cutting planes of the dual gradient for sensing.** Resonance frequency and Q-factor slices of $\Delta S = 0.15$ dual gradient. (a) Resonance frequency map of the dual gradient shown in Figure 4f-i. The seven horizontal dashed lines with a spacing of 10 pixels, indicate the cuts taken along the scaling direction of the dual gradient. (b) Cuts of the map shown in (a) with the resonance wavelength plotted against the x-position. All seven plots show linear and uniform resonance wavelength shifts across the gradient, almost independent of the y-position within the gradient. (c) Q-factor map of the dual gradient shown in Figure 4a-d. (d) Q-factor cut along the y-axis of the map shown in (c) where the cut is indicated as a dashed line. A clear trend towards lower Q-factors for increasing asymmetry (increasing y-position) is visible.

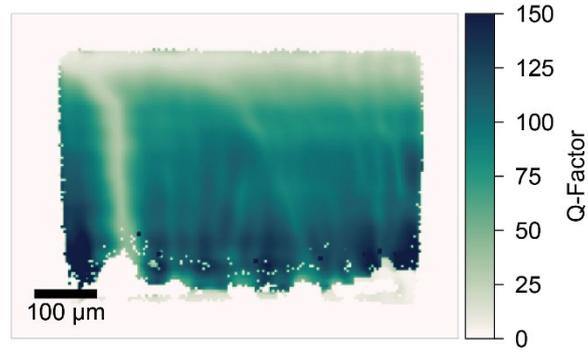

**Figure S14: Q-factor map of coated dual gradient.** Q-factor map of the dual-gradient metasurface of Figure 4e-i with an analyte concentration of 1%. The spectral fingerprint is clearly visible due to its strongly quenched Q-factors.

# Supplementary References